\documentclass[12pt]{article}
\usepackage{epsfig}
\usepackage{feynmp}
\usepackage{amsmath}

\newcommand{\bea}{\begin{eqnarray}}
\newcommand{\eea}{\end{eqnarray}}
\newcommand{\bd}{\begin{displaymath}}
\newcommand{\ed}{\end{displaymath}}
\newcommand{\be}{\begin{equation}}
\newcommand{\ee}{\end{equation}}

\renewcommand{\baselinestretch}{1.2}
\begin{document}

\thispagestyle{empty}

{\normalsize\sf
\rightline {hep-ph/0110282}
\rightline{IFT-01/29}
\vskip 3mm
\rm\rightline{October 2001}
}

\vskip 5mm

\begin{center}
  
{\LARGE\bf On suppressing the Higgsino-mediated proton decay in SUSY $\mathsf{SO(10)}$ GUT's.}

\vskip 10mm

\setcounter{footnote}{-1}

{\large\bf Krzysztof Turzy\'nski\footnote{email: \sl Krzysztof-Jan.Turzynski@fuw.edu.pl}}\\[5mm]

Institute of Theoretical Physics, Warsaw University\\
Ho\.za 69, 00-681 Warsaw, Poland

\end{center}

\vskip 5mm

\renewcommand{\baselinestretch}{1.1} 
\begin{abstract}
\vskip 3mm 
Using the freedom in $\mathsf{SO(10)}$ GUT's one can generalize the existing models without changing the mass spectrum of fermions to obtain a significant suppression of proton decay resulting from the dimension 5 operators with $\Delta B\neq 0$. In some limiting cases, these operators can be made negligible compared to dimension 6 operators resulting from the heavy gauge bosons exchange.
\end{abstract}
\renewcommand{\baselinestretch}{1.2}

\newpage \setcounter{page}{1} \setcounter{footnote}{0}

\begin{fmffile}{ProtonGraphs}


\section{Introduction.}
\setcounter{equation}{0}

Proton decay is the most generic qualitative prediction of Grand Unified Theories (GUT's, see \cite{buras},\cite{fritzsch},\cite{georgi},\cite{pati}). Unfortunately, quantitative predictions are subject to several well-known uncertainties. Not only depend they on the choice of the GUT, but, for a given theory, serious uncertainties reside in various ambiguities in the values of the relevant parameters. In supersymmetric (SUSY) GUT's, the dominant contributions to proton decay amplitude usually result from the dimension 5, $\Delta B\neq 0$ operators (\cite{dimopoulos},\cite{ellis},\cite{sakai},\cite{weinberg}), describing effectively the exchange of a colour triplet with mass of the order of $M_{GUT}$:
\begin{center}
\begin{fmfgraph*}(150,90)
\fmfleft{l1,l2}
\fmfright{r1,r2}
\fmf{fermion,label=$q_a$}{l1,v1}
\fmf{dashes_arrow,label=$\tilde{q}_b$}{l2,v1}
\fmf{fermion,label=$T$}{v2,v1}
\fmfv{label=$\mathcal{Y}_{ab}$}{v1}
\fmf{fermion,label=$q_c$}{r1,v2}
\fmf{dashes_arrow,label=$\tilde{l}_d$}{r2,v2}
\fmfv{label=$\tilde{\mathcal{Y}}_{cd}$}{v2}
\end{fmfgraph*}
\end{center}
The Wilson coefficients of the effective, dimension 5, $\Delta B\neq 0$ operators can be expressed in terms of the GUT parameters. The contribution of the diagram depicted above is:  
\begin{equation}
\label{toymodel}
C^{abcd} \sim \frac{1}{M_T} \mathcal{Y}_{ab} \tilde{\mathcal{Y}}_{cd}
\end{equation}
Since these coefficients are suppressed with only one power of the large mass scale\footnote{In the minimal $\mathsf{SU(5)}$ $M_T\sim M_{GUT}$ but numerous $\mathsf{SO(10)}$ models admit $M_T\sim M_\mathrm{Planck}$}, we can neglect the impact on proton decay of dimension 6 operators, induced by heavy gauge bosons exchange\footnote{If the heavy gauge boson exchange is the only source of proton decay, the proton lifetime $\tau\sim 10^{35}\mathrm{y}$; see \cite{hisano2}.}. The Wilson coefficients of the operators of dimension 5 have to be renormalized from the unification scale to the electroweak scale.  At the scale of SUSY breaking, which lies rather closely to the electroweak scale $M_Z$, the sparticles are integrated out and the operators of dimension 5 give rise to the effective four-fermion operators of dimension 6. This is depicted on the diagram below:   
\begin{center}
\begin{fmfgraph*}(200,130)
\fmfleft{l1,l2}
\fmfright{r1,r2}
\fmf{fermion,label=$q$}{l1,vert}
\fmf{fermion,label=$q$}{r1,vert}
\fmf{fermion,label=$q$}{l2,vl}
\fmf{fermion,label=$l$}{r2,vr}
\fmf{plain,tension=.5,label=$\tilde{w}\tilde{h}\tilde{g}$}{vl,vr}
\fmf{dashes_arrow,tension=.5,label=$\tilde{q}$}{vl,vert}
\fmf{dashes_arrow,tension=.5,label=$\tilde{l}$}{vr,vert}
\end{fmfgraph*} 
\end{center}
When the gauginos are much lighter than the sfermions, the Wilson coefficients $C'$ of dimension 6 operators can be approximated by: 
\begin{equation}
C' \sim \frac{g_2^2}{16\pi^2} \frac{m_{\tilde{w},\tilde{h}}}{m_{\tilde{q},\tilde{l}}^2} C
\end{equation}
Once the passage to the effective theory containing four-fermion operators is made, the renormalization procedure goes on to the scale of proton mass $\sim 1\,\mathrm{GeV}$. At this scale the link between three-quark operators and the interactions of hadrons is made using chiral Lagrangian technique, which exploits the approximate $\mathsf{SU(3)}_V\times\mathsf{SU(3)}_A$ flavour symmetry of the hadronic world (\cite{chadha}, \cite{claudson}). In this manner, proton lifetime can eventually be calculated.

The first step (i.e. calculating the Wilson coefficients of the dimension 5 operators with $\Delta B\neq 0$ in a specific SUSY GUT) is relatively unambiguous in the simplest $\mathsf{SU(5)}$ model, which was widely investigated and served as a laboratory for developing the remaining tools. However, the latest experimental lower bound on the proton lifetime in the dominant channel (\cite{sk}):
\begin{equation}
\label{experiment}
\tau \left( p\to K^+\bar{\nu} \right) > 6.7 \cdot 10^{32} \mathrm{y} \,\,\,\mathrm{at}\,\,\,90\%\,\mathrm{CL}
\end{equation}
indicates that the minimal $\mathsf{SU(5)}$ is ruled out (cf. \cite{murayama} for an updated discussion). Therefore, recently the attention has been focused on $\mathsf{SO(10)}$. Unlike the minimal $\mathsf{SU(5)}$, in which the Wilson coefficients of the $d=5$, $\Delta B\neq 0$ operators are determined by the Yukawa couplings and $M_T$ up to two phases\footnote{The impact of these phases on proton decay has been studied in \cite{goto}.}, $\mathsf{SO(10)}$ models are less constrained and thus less predicitive, but they give some hope to increase the prediction for proton lifetime in a natural way. Recently, several results in $\mathsf{SO(10)}$ have been reported (\cite{babu}, \cite{raby}), claiming maximal $\tau_p$ to be within one order of magnitude from the present experimental bounds.

The purpose of this paper is to take a critical look at those predictions. We shall use some straightforward generalizations of the model presented in \cite{babu} to illustrate some features of $\mathsf{SO(10)}$ GUT's that allow to suppress proton decay induced by the operators of dimension 5. In Section 2 we discuss thoroughly the contributions to the Wilson coefficients of the operators of dimension 5 relevant for proton decay and compare them with the contributions to Yukawa couplings. In Section 3 we stress the importance of the proper renormalization procedure and point out some subtleties that led to errors in the published works. In Section 4 three strategies for increasing the prediction for the proton lifetime are proposed. In Section 5 we explain why in $\mathsf{SO(10)}$ GUT's one usually needs large mass scales of the order of the Planck mass, which does not happen in case of the minimal $\mathsf{SU(5)}$, and sketch another method for raising the proton lifetime. The conclusions are presented in Section 6.


\section{The link between fermion masses and proton decay.}
\setcounter{equation}{0}

In this section we analyze a class of $\mathsf{SO(10)}$ GUT's which can provide a realistic mass spectrum for fermions\footnote{See \cite{babu} for an example of a theory of fermion masses.}. In these theories  the light Higgs doublets may reside in the vector representation $\mathbf{10_H}$ and in the spinor representation $\mathbf{16_H}$, i.e.:
\begin{eqnarray}
H_u &=& \mathbf{2}^{(u)}_{\mathbf{10_H}} \nonumber \\
\label{mixture}
H_d &=& \cos\gamma_d \mathbf{2}^{(d)}_{\mathbf{10_H}}+\sin\gamma_d \mathbf{2}^{(d)}_{\mathbf{16_H}}
\end{eqnarray}
We also assume that even at the GUT scale we are dealing with an effective theory resulting from Planck-scale physics which admits:
\begin{itemize}
\item renormalizable interactions described by operators consisting of three superfields (Yukawa couplings)
\item non-renormalizable interactions described by operators consisting of four superfields\footnote{Operators with more vevs have also been considered in the literature, but in a different context (Cf. \cite{raby}). We make a heuristic assumption that the effective couplings containing more vevs of the order of the unification scale are suppressed by higher powers of the quantity $\frac{M_{GUT}}{M_\mathrm{Planck}}$.}. They become effective Yukawa couplings, when one of the fields acquires a vacuum expectation value (vev) of the order of the unification scale. 
\end{itemize}
We also assume an additional global $\mathbf{Z_2}$ discrete symmetry, under which $\mathbf{16}_a$ are odd for $a=1,2,3$ and all the remaining fields are even (which is called matter parity).

We shall use the notation in which the symbol $[\ldots ]_R[\ldots ]$ signifies that the representations in the left and the right square brackets are contracted to form a representation $R$ (or $R+\bar{R}$, if necessary), respectively. In this notation all the effective Yukawa couplings which are consistent with the assumed symmetries read:
\begin{eqnarray}
\label{bjeden}
& \mathbf{16}_a\,\mathbf{16}_b\,\mathbf{10_H} & \\
\label{bdwa}
& \left[ \mathbf{16}_a\,\mathbf{16}_b\right]_{\mathbf{10}}\left[\langle \mathbf{45_H} \rangle \mathbf{10_H}\right] & \\
\label{btrzy}
& \left[ \mathbf{16}_a\,\mathbf{16}_b\right]_{\mathbf{120}}\left[\langle \mathbf{45_H} \rangle \mathbf{10_H}\right] & \\
\label{bcztery}
& \left[ \mathbf{16}_a\langle\mathbf{16_H}\rangle\right]_{\mathbf{10}}\left[\mathbf{16}_b\, \mathbf{16_H}\right] & \\
\label{bpiec}
& \left[ \mathbf{16}_a\langle\mathbf{16_H}\rangle\right]_{\mathbf{120}}\left[\mathbf{16}_b\, \mathbf{16_H}\right] & \\
\label{bszesc}
& \left[ \mathbf{16}_a\langle\mathbf{\bar{16}_H}\rangle\right]_{\mathbf{1}}\left[\mathbf{16}_b\, \mathbf{\bar{16}_H}\right] & \\
\label{bsiedem}
& \left[ \mathbf{16}_a\langle\mathbf{\bar{16}_H}\rangle\right]_{\mathbf{45}}\left[\mathbf{16}_b\, \mathbf{\bar{16}_H}\right] & \\
\label{bosiem}
& \left[ \mathbf{16}_a\langle\mathbf{\bar{16}_H}\rangle\right]_{\mathbf{210}}\left[\mathbf{16}_b\, \mathbf{\bar{16}_H}\right] & 
\end{eqnarray}


The symmetry group $\mathsf{SO(10)}$ can be broken to the Standard Model gauge group $G_{SM}=\mathsf{SU(3)}_C\times\mathsf{SU(2)}_L\times\mathsf{U(1)}_Y$ by the vevs of the adjoint and spinor Higgs fields:
\begin{eqnarray}
\langle \mathbf{45_H} \rangle &\sim & B-L \nonumber\\
\label{bzero}
\langle \mathbf{16_H} \rangle &\sim & \nu^c
\end{eqnarray}
(the conjugate spinor representation $\mathbf{\bar{16}_H}$ must also acquire a vev in SUSY theories). Then the coupling (\ref{btrzy}) introduces mass differences between the down-type quarks and leptons, the coupling (\ref{bcztery}) and (\ref{bpiec}) provide a non-trivial Kobayashi-Maskawa matrix and the couplings (\ref{bszesc})--(\ref{bosiem}) generate Majorana masses for right-handed neutrinos when both the spinor representations $\mathbf{\bar{16}_H}$ acquire vevs.

Now we give the explicit formulae for the Yukawa couplings and the couplings to colour triplets resulting from the operators (\ref{bjeden})--(\ref{bosiem}). We denote by $T_R$ ($\bar{T}_R$) a triplet (antitriplet) coming from the representation $R$. After the GUT symmetry is broken, the part of the superpotential resulting from these operators reads\footnote{We use Greek letters for colour indices, lower Latin letters $i$-$r$ for weak isospin indices, lower Latin letters $a$-$e$ and upper Latin letters for flavour indices.}:
\begin{eqnarray}
W_Y &=& Y^U_{ab} \hat{U}^\mathsf{c}_{a\alpha} \hat{U}^\alpha_b \hat{H_u} + Y^D_{ab} \hat{D}^\mathsf{c}_{a\alpha} \hat{D}^\alpha_b \hat{H_d} + Y^E_{ab} \hat{E}^\mathsf{c}_a \hat{E}_b \hat{H_d} + \nonumber\\
& & + \mathcal{Y}^{(1)}_{ab} \hat{U}^\alpha_a \hat{D}^\beta_b \hat{T}^\gamma_\mathbf{10} \varepsilon_{\alpha\beta\gamma}+ \mathcal{Y}^{(2)}_{ab} \hat{U}^\alpha_a \hat{D}^\beta_b \hat{T}^\gamma_\mathbf{\bar{16}} \varepsilon_{\alpha\beta\gamma} + \nonumber\\
& & + \mathcal{X}^{(1)}_{ab} \hat{U}^\mathsf{c}_{a\alpha} \hat{D}^\mathsf{c}_{b\beta} \hat{\bar{T}}_{\mathbf{10}\gamma} \varepsilon^{\alpha\beta\gamma} + \mathcal{X}^{(2)}_{ab} \hat{U}^\mathsf{c}_{a\alpha} \hat{D}^\mathsf{c}_{b\beta} \hat{\bar{T}}_{\mathbf{16}\gamma} \varepsilon^{\alpha\beta\gamma} + \nonumber\\
& & + \tilde{\mathcal{Y}}^{(1)}_{ab} \left( \hat{U}^\alpha_a \hat{E}_b - \hat{D}^\alpha_a \hat{\nu}_b \right) \hat{\bar{T}}_{\mathbf{10}\alpha} + \tilde{\mathcal{Y}}^{(2)}_{ab} \left( \hat{U}^\alpha_a \hat{E}_b - \hat{D}^\alpha_a \hat{\nu}_b \right) \hat{\bar{T}}_{\mathbf{16}\alpha} + \nonumber\\
& & + \tilde{\mathcal{X}}^{(1)}_{ab} \hat{U}^\mathsf{c}_{a\alpha} \hat{E}^\mathsf{c}_b \hat{T}^\alpha_\mathbf{10} + \tilde{\mathcal{X}}^{(2)}_{ab} \hat{U}^\mathsf{c}_{a\alpha} \hat{E}^\mathsf{c}_b \hat{T}^\alpha_\mathbf{\bar{16}} + \nonumber\\
& & + M^R_{ab} \hat{\nu}^\mathsf{c}_a \hat{\nu}^\mathsf{c}_b
\end{eqnarray}
The matrices $Y^U$, $Y^D$, $Y^E$ are the Yukawa couplings of the up-type quarks, down-type quarks and leptons, respectively. The Majorana mass matrix of the right-handed neutrinos is denoted by $M^R$. The matrices $\mathcal{Y}^{(i)}$ and $\tilde{\mathcal{Y}}^{(i)}$ ($\mathcal{X}^{(i)}$ and $\tilde{\mathcal{X}}^{(i)}$) are the Yukawa-like couplings of the left-handed (right-handed) matter fields and colour triplets and antitriplets, respectively. If we denote by $a^{(n)}_{ij}$ the contributions resulting from eq. (\ref{bzero}$+n$), all the abovementioned matrices can be expressed in the following way:
\begin{eqnarray}
\label{pocztable2}
Y^U_{ab} &=& a^{(1)}_{(ab)} + a^{(3)}_{[ab]} \\
Y^D_{ab} &=& -\left( a^{(1)}_{(ab)} + a^{(3)}_{[ab]} \right) \cos\gamma_d + \left( a^{(4)}_{(ab)} + a^{(5)}_{(ab)} \right) \sin\gamma_d \\
Y^L_{ab} &=& -\left( a^{(1)}_{(ab)} -3 a^{(3)}_{[ab]} \right) \cos\gamma_d + \left( a^{(4)}_{(ab)} + a^{(5)}_{(ab)} \right) \sin\gamma_d \\
\mathcal{Y}^{(1)}_{ab} &=& a^{(1)}_{(ab)} + a^{(2)}_{(ab)} \\
\mathcal{Y}^{(2)}_{ab} &=& a^{(7)}_{(ab)} + a^{(8)}_{(ab)} \\
\tilde{\mathcal{Y}}^{(1)}_{ab} &=& a^{(1)}_{(ab)} - a^{(2)}_{(ab)} -2 a^{(3)}_{[ab]} \\
\tilde{\mathcal{Y}}^{(2)}_{ab} &=& a^{(4)}_{(ab)} + a^{(5)}_{(ab)} \\
\mathcal{X}^{(1)}_{ab} &=& a^{(1)}_{(ab)} - a^{(2)}_{(ab)} \\
\mathcal{X}^{(2)}_{ab} &=& a^{(4)}_{(ab)} + a^{(5)}_{(ab)} \\
\tilde{\mathcal{X}}^{(1)}_{ab} &=& a^{(1)}_{(ab)} + a^{(2)}_{(ab)} +2 a^{(3)}_{[ab]} \\
\tilde{\mathcal{X}}^{(2)}_{ab} &=& a^{(7)}_{(ab)} + a^{(8)}_{(ab)} \\
\label{koniectable2}
M^R_{ab} &=& \left( a^{(6)}_{(ab)} + \frac{5}{4} \left(a^{(7)}_{(ab)} + a^{(8)}_{(ab)} \right) \right) M_U
\end{eqnarray}
where $M_U=|\langle \mathbf{\bar{16}_H} \rangle|$.

Having integrated out heavy triplets, we obtain the following effective Lagrangian.
\begin{equation}
\mathcal{L} = \mathcal{L}_{MSSM} + C^{abcd}\mathcal{O}_{abcd}+ \tilde{C}^{abcd}\tilde{\mathcal{O}}_{abcd}
\end{equation}
where:
\begin{eqnarray}
\label{operatory}
\mathcal{O}_{abcd} &=& \left[ \hat{Q}^\alpha_{ai}\hat{Q}^\beta_{bj}\hat{Q}^\gamma_{ck}\hat{L}_{dl} \right]_F \varepsilon_{ij}\varepsilon_{kl}\varepsilon_{\alpha\beta\gamma } \\
\tilde{\mathcal{O}}_{abcd} &=& \left[ \hat{U}^\mathsf{c}_{a\alpha }\hat{D}^\mathsf{c}_{b\beta }\hat{U}^\mathsf{c}_{c\gamma }\hat{E}^\mathsf{c}_{d} \right]_F \varepsilon^{\alpha\beta\gamma } 
\end{eqnarray}
We shall call $\mathcal{O}_{abcd}$ and $\tilde{\mathcal{O}}_{abcd}$ the $LLLL$ and $RRRR$ operators.

We denote by $M^{-1}$ the inverse of the mass matrix for the colour triplets. The elements of this matrix are labelled by the relevant representations of $\mathsf{SO(10)}$ to which the triplets belong. Using the equations (\ref{pocztable2})--(\ref{koniectable2}) we obtain:
\begin{eqnarray}
2C^{abcd} &=& \left( a^{(1)}_{(ab)} + a^{(2)}_{(ab)} \right) \left(  a^{(1)}_{(cd)} - a^{(2)}_{(cd)} -2 a^{(3)}_{[cd]} \right) M^{-1}_{\mathbf{10}\,\mathbf{10}} +  \nonumber\\
& &  + \left( a^{(7)}_{(ab)} + a^{(8)}_{(ab)} \right) \left(  a^{(1)}_{(cd)} - a^{(2)}_{(cd)} -2 a^{(3)}_{[cd]} \right) M^{-1}_{\mathbf{10}\,\mathbf{16}} + \nonumber\\
& &  +  \left( a^{(1)}_{(ab)} + a^{(2)}_{(ab)} \right) \left(  a^{(4)}_{(cd)} + a^{(5)}_{(cd)} \right) M^{-1}_{\mathbf{16}\,\mathbf{10}} +  \nonumber\\
\label{matka1}
& &  +  \left( a^{(7)}_{(ab)} + a^{(8)}_{(ab)} \right) \left(  a^{(4)}_{(cd)} + a^{(5)}_{(cd)} \right) M^{-1}_{\mathbf{16}\,\mathbf{16}} \\
& & \nonumber\\
 \tilde{C}^{abcd} &=& \left( a^{(1)}_{(ab)} - a^{(2)}_{(ab)} \right) \left(  a^{(1)}_{(cd)} + a^{(2)}_{(cd)} +2 a^{(3)}_{[cd]} \right) M^{-1}_{\mathbf{10}\,\mathbf{10}} +  \nonumber\\
& &  + \left( a^{(4)}_{(ab)} + a^{(5)}_{(ab)} \right) \left(  a^{(1)}_{(cd)} + a^{(2)}_{(cd)} +2 a^{(3)}_{[cd]} \right) M^{-1}_{\mathbf{10}\,\mathbf{16}} + \nonumber\\
& &  +  \left( a^{(1)}_{(ab)} - a^{(2)}_{(ab)} \right) \left(  a^{(7)}_{(cd)} +a^{(8)}_{(cd)} \right) M^{-1}_{\mathbf{16}\,\mathbf{10}} +  \nonumber\\
\label{matka2}
& &  +  \left( a^{(4)}_{(ab)} + a^{(5)}_{(ab)} \right) \left(  a^{(7)}_{(cd)} + a^{(8)}_{(cd)} \right) M^{-1}_{\mathbf{16}\,\mathbf{16}}
\end{eqnarray}
Factor $2$ results from the following identity:
\begin{equation}
\hat{Q}^\alpha_{ai} \hat{Q}^\beta_{bj} \varepsilon_{ij} \varepsilon_{\alpha\beta\gamma} = 2 \hat{U}^\alpha_{\left( a \right. } \hat{D}^\beta_{\left. b \right)}  \varepsilon_{\alpha\beta\gamma}
\end{equation}
In the doublet-triplet splitting mechanism presented in \cite{babu}, $M^{-1}_{\mathbf{10}\,\mathbf{16}}=0$ in order to keep one pair of the weak doublets massless at the unification scale. The remaining Higgs fields, including the colour triplets, have masses of the order of $M_{GUT}$ then.

\subsection{The freedom of a general model.}

The coupling (\ref{bdwa}) does not contribute to the Yukawa couplings but affects proton decay. Indeed, when the gauge symmetry $\mathsf{SO(10})$ is broken to $G_{SM}$, the vector representation $\mathbf{10_H}$ splits into two weak doublets which carry no $B-L$ charge and two colour triplets with $B-L=\pm\frac{2}{3}$. Because $\langle \mathbf{45_H} \rangle \sim B-L$ is evaluated directly on this vector representation, only the couplings to colour triplets can arise from (\ref{bdwa}).

The coupling (\ref{bszesc}) contributes only to the Majorana masses of the right-handed neutrinos and has no impact on proton decay. This can be seen as follows. The vev $\langle \mathbf{\bar{16}_H} \rangle$ is invariant under $G_{SM}$ and it is contracted with some matter field to give a $\mathsf{SO(10)}$ singlet, which is, of course, a $G_{SM}$ singlet. Thus, this matter field must be the only $G_{SM}$ singlet in the matter multiplet: $\nu^c$. Because both the right-handed neutrinos and the colour triplets decouple near the unification scale, the coupling (\ref{bszesc}) cannot contribute to the dimension 5 operators with $\Delta B\neq 0$. Since the combinations of the couplings couplings (\ref{bszesc})--(\ref{bosiem}) that enter the Majorana masses of the right-handed neutrinos and those enetring the amplitudes of proton decay are linearly independent, there is no generic link between the Majorana masses of the right-handed neutrinos and proton decay.

\subsection{Babu-Pati-Wilczek model.}

In \cite{babu}, a quantitative GUT of the type described above is presented. A realistic spectrum of fermion masses can be obtained for $a^{(1)}_{33}=1$, $a^{(1)}_{23}=-0.1096$, $a^{(1)}_{11}=10^{-5}$, $-a^{(2)}_{23}=a^{(3)}_{23}=-0.0954$, $a^{(4)}_{23}+a^{(5)}_{23}=-0.0413$, $a^{(4)}_{12}+a^{(5)}_{12}=4.14\cdot 10^{-3}$ and the remaining entries in the matrices $a^{(1)},\ldots,a^{(5)}$ equal 0. Some assumptions about the spectrum of the theory at the Planck scale (e.g. the couplings (\ref{bdwa}) and (\ref{btrzy}) can only arise due to an exchange of $\mathbf{16}+\mathbf{\bar{16}}$ states with masses of the Planck scale, the matrices $a^{(6)},\ldots,a^{(8)}$ are chosen so that the contributions to the Yukawa couplings of neutrinos are equal to that to proton decay) remove all the freedom discussed above. This arbitrary choice makes the model maximally predictive, but the predictions for the proton lifetime are inconsistent with the experimental bound (\ref{experiment}). Therefore, we shall treat this model a starting point for some generalizations which allow to increase the predictions for the proton lifetime.


\section{The importance of the proper renormalization.}
\setcounter{equation}{0}
 
In the minimal $\mathsf{SU(5)}$ model, Wilson coefficients of the $LLLL$ operators, which result from integrating out heavy colour triplets $T$, can be written as \cite{hisano}:
\begin{equation}
\label{wilsonsu5}
C^{abcd} = \frac{1}{M_T} \mathcal{Y}^{ab}\tilde{\mathcal{Y}}^{cd}
\end{equation}
where:
\begin{eqnarray}
\label{eq54}
\mathcal{Y}(M_{GUT}) &=& Y^U(M_{GUT}) \\
\label{eq55}
\tilde{\mathcal{Y}}(M_{GUT}) &=& Y^D(M_{GUT}) = Y^E(M_{GUT})
\end{eqnarray}
Though the running of $\mathcal{Y}$ and $\tilde{\mathcal{Y}}$ differs from the running of the Yukawa matrices and the equalities (\ref{eq54}) and (\ref{eq55}) do not hold except at the unification scale, they can help to connect the observables from the Standard Model with the Wilson coefficients relevant for proton decay. The following two recipes can be applied.

{\bf Recipe 1.} Within the MSSM the Yukawa couplings $y_c(M_Z)$, $y_d(M_Z)$, etc. can be expressed by $m_c(1\mathrm{GeV})$, $m_d(1\mathrm{GeV})$, etc  using the renormalization group (RG) analysis. Next, Yukawa couplings are evolved upwards to $M_{GUT}$, where $\mathcal{Y}$ and $\tilde{\mathcal{Y}}$ are calculated by using the equations (\ref{eq54}) and (\ref{eq55}). At the GUT scale we use (\ref{wilsonsu5}) to calculate Wilson coefficients, which are then renormalized downwards to $M_Z$.
If $a,b,c,d=1,2$, the renormalization group equations (RGE's) for $C^{abcd}$ and $\tilde{C}^{abcd}$ can easily be solved by using the 1-loop solutions for the running gauge couplings. We obtain the following enhancement factors (\cite{hisano}):
\begin{eqnarray}
\tilde{A}_S^{abcd} &=& \frac{C^{abcd}(M_Z)}{C^{abcd}(M_{GUT})} = \nonumber\\
\label{wzmwilson1}
&=& \left( \frac{g_3^2(M_Z)}{g_3^2(M_{GUT})} \right)^{\frac{4}{3}} \left( \frac{g_2^2(M_Z)}{g_2^2(M_{GUT})} \right)^{-3} \left( \frac{g_1^2(M_Z)}{g_1^2(M_{GUT})} \right)^{-\frac{1}{33}} \\
\tilde{A}_S^{'abcd} &=& \frac{\tilde{C}^{abcd}(M_Z)}{\tilde{C}^{abcd}(M_{GUT})}= \nonumber\\
\label{wzmwilson2}
&=& \left( \frac{g_3^2(M_Z)}{g_3^2(M_{GUT})} \right)^{\frac{4}{3}} \left( \frac{g_1^2(M_Z)}{g_1^2(M_{GUT})} \right)^{-\frac{2}{11}}
\end{eqnarray}

The RGE's for the Wilson coefficients of the operators involving the third generation of fermions must be solved numerically, since the running is affected by the large Yukawa couplings. At $M_Z$ all SUSY particles are decoupled, and dimension 5 operators give rise to dimension 6 operators with coefficients $C^{'abcd}$, which are renormalized downwards to $1\,\mathrm{GeV}$. This results in another enhancement factor:
\begin{equation}
\label{also10}
\tilde{A}_L =  \frac{C^{'abcd}(1\,\mathrm{GeV})}{C^{'abcd}(M_Z)}\sim 1.35
\end{equation}
Then the amplitudes of proton decay can eventually be calculated.

{\bf Recipe 2.} We can also use the equations (\ref{eq54}) and (\ref{eq55}) to define the quantity:
\begin{equation}
C_\star^{abcd}(t) = \frac{1}{M_T} Y^U_{ab}(t) Y^D_{cd}(t)
\end{equation}
which obeys:
\begin{equation}
C_\star^{abcd}(M_{GUT}) = C^{abcd}(M_{GUT})
\end{equation}
This is not the Wilson coefficient relevant for proton decay, because it is renormalized in a different way. However, knowing the running of the Yukawa couplings and the Wilson coefficients, we can calculate the discrepancy between $C_\star^{abcd}(M_Z)$ and $C^{abcd}(M_Z)$. This is the suppression factor $A_S$ introduced in ref. \cite{ellis} and calculated (for $m_t=100\,\mathrm{GeV}$) in ref. \cite{hisano}:
\begin{equation}
A_S = \left( \frac{g_3^2(M_Z)}{g_3^2(M_{GUT})} \right)^{-\frac{4}{9}} \left( \frac{g_1^2(M_Z)}{g_1^2(M_{GUT})} \right)^{\frac{7}{99}} \sim 0.67
\end{equation}
Then we can express $C_\star^{abcd}(t)$ in terms of the running masses. The coefficient of a prototype operator relevant for proton decay reads:
\begin{equation}
C^{abcd}(M_Z) = A_S C_\star^{abcd}(M_Z) = A_S\frac{g_2^2}{M_T M_W^2 \sin 2\beta_H} m_{u^a}(M_Z) m_{d^c}(M_Z) \delta^{ab} V_{CKM}^{cd*}
\end{equation}
since at the $M_Z$ scale all the Yukawa couplings can be expressed in terms of the running masses at that scale.
After dressing with superparticles we can repeat this procedure to dimension 6 operators. Between $M_Z$ and $1\,\mathrm{GeV}$ the discrepancy between the exact solution of the RGE for true Wilson coefficient and the coefficients expressed by running masses results in another suppression factor introduced in refs. \cite{ellis}, \cite{hisano}. Having solved the 1-loop RGE's for the strong coupling constant and the running masses, we obtain:  
\begin{equation}
A_L = \left( \frac{g_3^2(m_b)}{g_3^2(M_Z)} \right)^{-\frac{18}{23}} \left( \frac{g_3^2(m_c)}{g_3^2(m_b)} \right)^{-\frac{18}{25}} \left( \frac{g_3^2(1\,\mathrm{GeV})}{g_3^2(m_c)} \right)^{-\frac{2}{3}} = 0.43
\end{equation}
and the following relation:
\begin{equation}
\label{relder}
\tilde{A}_L^{-3} = A_L
\end{equation}
It has been pointed out in ref. \cite{arafune} (cf. a recent discussion in \cite{dermisek}) that this value differs from that given in the abovementioned papers. Solution of the 3-loop RGE's for the strong coupling constant and the running masses implies:
\begin{equation}
A_L = 0.28
\end{equation}
The relation (\ref{relder}) is also modified then.

Note that the meaning of $\tilde{A}^{abcd}_S$ and $\tilde{A}_L$ is completely different from that of $A_S$ and $A_L$, respectively, since the former cover all the renormalization effects, not the discrepancy in running of two quantities. 
The Recipe 2 cannot be conveniently used in the $\mathsf{SO(10)}$ models, because, as we have demonstrated in Section 2, the Wilson coefficients relevant for proton decay cannot be parametrized with the Yukawa matrices in a simple manner. However, the coefficients $\tilde{A}^{abcd}_S$ and $\tilde{A}^{'abcd}_S$, which do not depend on the choice of GUT, can be calculated either in the minimal $\mathsf{SU(5)}$ or in $\mathsf{SO(10)}$. In the former case, they are functions of fermion masses at $M_Z$, $\tan\beta$ and the corresponding $A_S$ factor\footnote{In fact, $A_S$ depends on the generation indices of the relevant Wilson coefficient and on the mass of the top quark. This has been analyzed in \cite{hisano}. The value quoted here was obtained for $m_t=100\,\mathrm{GeV}$.}.  

In particular, it is not legitimate to use the values of $A_S$ and $A_L$ given in ref. \cite{hisano} to account for all the RG effects in $\mathsf{SO(10)}$ models. However, we have found that in some of the previous works on $\mathsf{SO(10)}$ models these renormalizations were computed incorrectly. For example, the authors of ref. \cite{babu} use the wrong value $A_S=0.67$ and obtain the upper theoretical bound on the proton lifetime at least twice too large.


\section{How to increase the prediction for the proton lifetime?}
\setcounter{equation}{0}

In order to make definite predictions for the proton lifetime we assume that Majorana masses of the right-handed neutrinos are generated solely by the coupling (\ref{bszesc}) and $a^{(7)}=a^{(8)}=0$, except for Section \ref{numasses}, in which the possible link between the Majorana masses of the right-handed neutrinos and proton decay is investigated. We consider the following two SUSY spectra: 
\begin{itemize}
\item all sfermions have masses $1000\,\mathrm{GeV}$ except the lighter stop, whose mass is taken to be $400\,\mathrm{GeV}$, the masses of charginos are $100\,\mathrm{GeV}$ and $500\,\mathrm{GeV}$, gluinos have masses $350\,\mathrm{GeV}$\item inspired by \cite{murayama}, we also analyzed the case in which the first two generations of all sfermions have masses $10\,\mathrm{TeV}$, the third generation has masses $1\,\mathrm{TeV}$ except the lighter stop with mass $400\,\mathrm{GeV}$; we shall call this case the decoupling.
\end{itemize}
We neglect the contributions of the neutralinos to proton decay. We also use the latest results of the JLQCD collaboration for the hadronic matrix element $|\alpha |=|\beta |=0.015\,\mathrm{GeV}^3$ (\cite{aoki}). The decay channels are enumerated in the following way:
\begin{center}
\begin{tabular}{|c|l|}
\hline
No. & decay channel \\
\hline
1 & $p\to K^+\bar{\nu}$ \\
2 & $p\to \pi^+\bar{\nu}$ \\
3 & $p\to K^0\mu^+$ \\
4 & $p\to \pi^0\mu^+$ \\
\hline
\end{tabular}
\end{center}

\subsection{Neutrino masses and proton decay.}
\label{numasses}

\begin{figure}
\begin{center}
\rotatebox{0}{
\epsfig{file=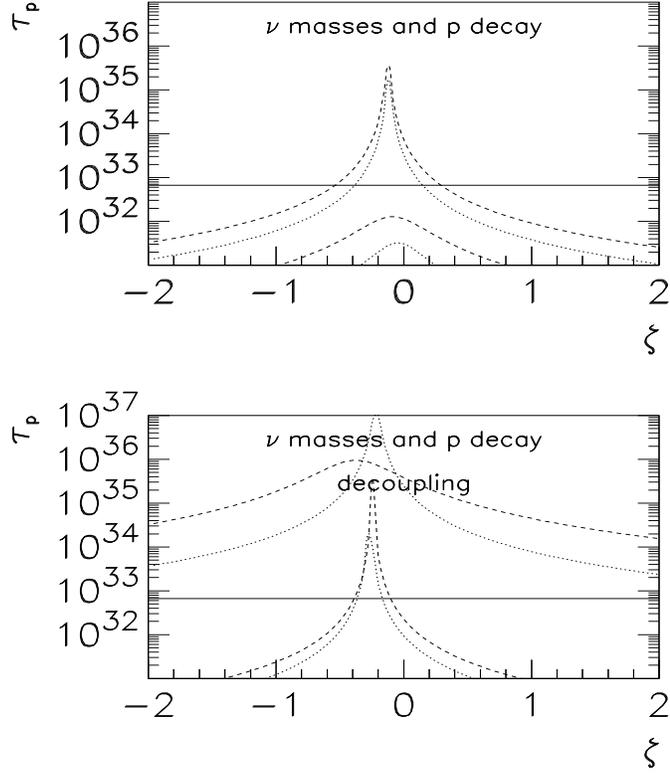,height=15cm,width=12cm}}
\caption{The proton lifetime (in years) in the dominant channels as a function of $\zeta$. Only the $\Delta B\neq 0$ operators of dimension 5 have been taken into account. The lower dotted line, the lower dashed line, the upper dotted line and the upper dashed line are the partial proton lifetimes in the channels 1,2,3 and 4, respectively. The solid line represents the current experimental lower bound on the proton lifetime in the decay channel 1. See Section \ref{numasses} for more explanation.\label{figa5}}
\end{center}
\end{figure}

In order to study he impact of $a^{(7)}$ and $a^{(8)}$ on proton decay we assume that $a^{(7)}+a^{(8)}$ is the fraction of the Majorana mass matrix of the right-handed neutrinos normalized to $M_U$:
\begin{equation}
\label{rhneutrina}
a^{(7)} + a^{(8)} = \frac{\zeta}{M_U} M_R
\end{equation}
We also assume $a^{(2)}=a^{(3)}$. In the limit of $\zeta\to 0$, the model becomes what is called the Strategy I with $\xi=1$ in the following section. The prediction for the proton lifetime as a function of $\zeta$ for the two SUSY spectra is depicted on Figure \ref{figa5}. The maximum is shifted towards $\zeta<0$ due to a destructive interference of the contributions resulting from (\ref{bsiedem}) and (\ref{bosiem}) with the remaining ones.

\subsection{Strategy I.}
\label{s42}

In \cite{babu}, in the model dubbed Case II, the coupling of the form
\begin{equation}
\label{b23}
\left[ \mathbf{16}_2 \mathbf{10_H} \right]_{\mathbf{16}+\mathbf{\bar{16}}} \left[ \langle \mathbf{45_H} \rangle \mathbf{16}_3 \right]
\end{equation}
is used to provide flavour-antisymmetric contributions to the Yukawa couplings which allow the down-type quarks and leptons of a given generation have different masses. The operator (\ref{b23}) is, of course, a combination of the operators (\ref{bdwa}) and (\ref{btrzy}). This can also be obtained with the use of the coupling:
\begin{equation}
\label{tildeb23}
- \left[ \mathbf{16}_3 \mathbf{10_H} \right]_{\mathbf{16}+\mathbf{\bar{16}}} \left[ \langle \mathbf{45_H} \rangle \mathbf{16}_2 \right]
\end{equation}
Then the operators (\ref{b23}) and (\ref{tildeb23}) give the same contributions to $a^{(3)}$, whereas their contributions to $a^{(2)}$ have opposite signs. In general, both of them can be present in the theory. Then $a^{(2)}$ and $a^{(3)}$ are no longer equal but proportional:
\begin{equation}
a^{(2)} = \xi a^{(3)}
\end{equation}
The dependence of the proton lifetime on the parameter $\xi$ is depicted on Figures \ref{figa2}.



There are three special cases $\xi=-1,0,1$, in which the sum of the operators (\ref{bdwa}) and (\ref{btrzy}) can be represented by a single $\mathsf{SO(10)}$ invariant operator. For $\xi=1$ there is an accidental cancellation of the various contributions to the $LLLL$ operators. This is due to the fact that, numerically, $\mathcal{Y}={Y^U}^\dagger Y^U$ and it is diagonalized simultaneously with the Yukawa couplings. The maximum is shifted to $\xi>1$ because of the interference with the $RRRR$ operators. In the case of the decoupling the proton decay results mainly from the stop exchange in the SUSY loop and the coupling to the third generation is quite insensitive to changing $\xi$.


\subsection{Strategy II.}
\label{s43}

It follows from (\ref{matka1}) that if
\begin{equation}
\label{finetune}
a^{(2)}=-a^{(1)}
\end{equation}
the Wilson coefficients $C^{abcd}$ are driven to zero and only the operators consisting of right-handed superfields are relevant for proton decay. They are, however, strongly suppressed by the smallness of the Yukawa couplings of the first two generations of fermions. 

\begin{figure}
\begin{center}
\rotatebox{0}{
\epsfig{file=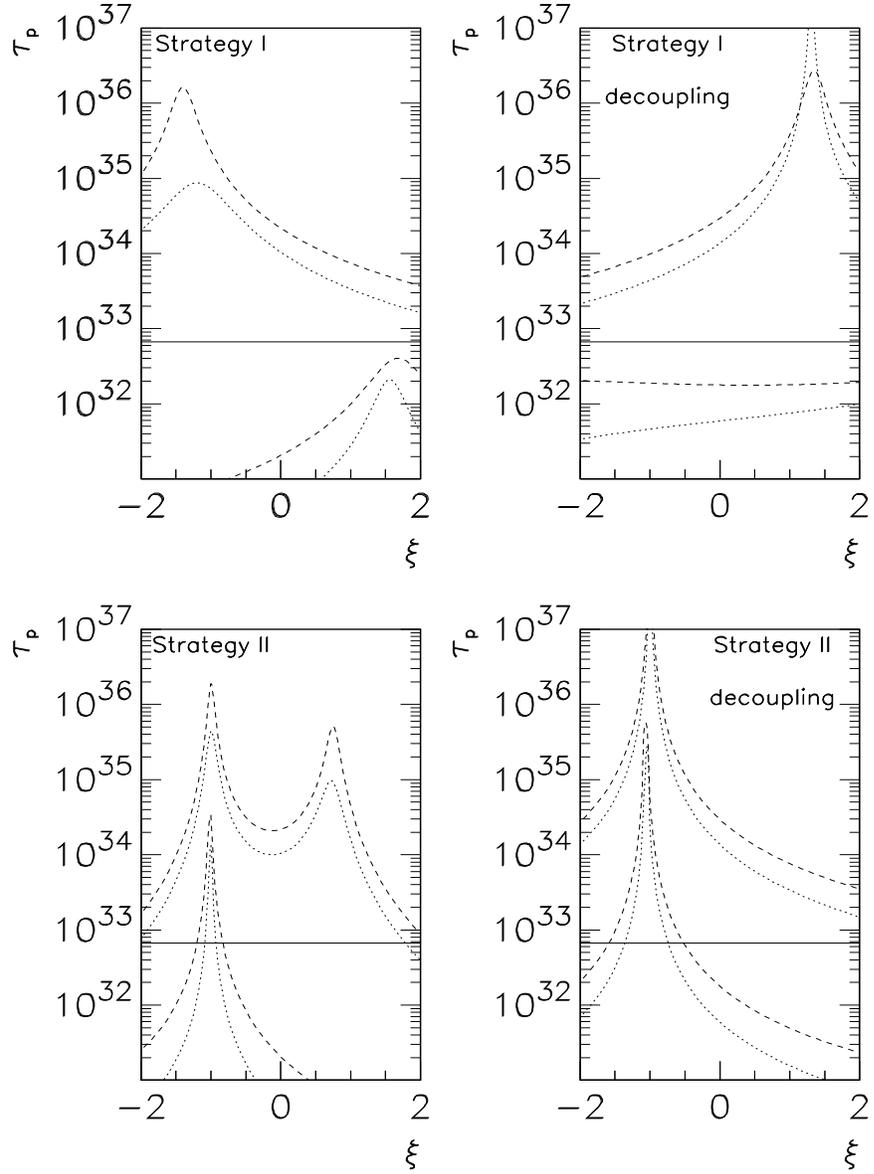,height=20cm,width=15cm}}
\caption{The proton lifetime (in years) in the dominant channels as a function of $\xi$ for Strategies I and II. See the caption of Figure (\ref{figa5}) and Sections \ref{s42} and \ref{s43} for the explanation.\label{figa2}}
\end{center}
\end{figure}

One could however argue that the assumption (\ref{finetune}) is an extreme fine-tuning. Therefore, we shall consider a slightly more general scenario, in which:
\begin{eqnarray}
a^{(2)} = \xi a^{(1)}
\end{eqnarray}
The dependence of the proton lifetime on the parameter $\xi$ is depicted on Figure \ref{figa2}.

For $\xi=-1$ there is a maximum of the lifetime in all the channels, because the contribution of the $LLLL$ operators vanish and the proton decay is induced solely by the $RRRR$ operators. The maximum at $\xi=1$ for the channels $p\to K^0\mu^+$ and $p\to \pi^0\mu^+$ results from vanishing the coefficients of the $RRRR$ operators.

The agreement with the experimental data can only be obtained in a narrow band around $\xi=-1$. Though this solution is slightly more natural in the case of decoupling, it still requires some fine-tuning of the parameters.


\subsection{Strategy III.}
\label{s44}

This idea utilizes an additional adjoint representation $\mathbf{45_R}$, which acquires a vev proportional to the right isospin of the Pati-Salam model \cite{pati}. Then the coupling (\ref{bjeden}) can be replaced with\footnote{This pushes the value of $V_{cb}$ slightly out of the experimentally allowed region. However, we regard this strategy only an illustration of a general possibility of model building.}:
\begin{equation}
\label{45}
\left[ \mathbf{16}_a \mathbf{10_H} \right]_{\mathbf{16}+\mathbf{\bar{16}}} \left[ \langle \mathbf{45_R} \rangle \mathbf{16}_a \right]
\end{equation}
Because the right isospins of the left-handed particles equal zero, this coupling does not contribute to $C^{abcd}$. Moreover, the contribution to $\tilde{C}^{abcd}$ is antisymmetric in the first two indices, since the right isospin of the up-type antiquarks is minus that of the down-type antiquarks.

We assume   
\begin{equation}
a^{(2)} = \xi a^{(3)}
\end{equation}  
The dependence of the proton lifetime on the parameter $\xi$ is depicted on Figures \ref{figa4}.



The maximum of the proton lifetime in all the dominant channels at $\xi=0$ results from vanishing the coefficients of the $LLLL$ operators there and the prediction for the proton lifetime is consistent with the expreimental data. The case of $\xi=0$ is equivalent to the absence of the coupling (\ref{bdwa}), so it satisfies the requirement of naturalness.


\section{Why does one usually need larger $M_T$ in $\mathsf{SO(10)}$ models than in the minimal $\mathsf{SU(5)}$? Strategy IV.}
\setcounter{equation}{0}
\label{s5}

In the minimal $\mathsf{SU(5)}$ there are two Yukawa couplings: one for the up-type quarks and a common coupling for the down-type quarks and leptons. The model predicts that at the unification scale the Yukawa couplings are equal to the relevant couplings to the colour triplets $\mathcal{Y}$ and $\tilde{\mathcal{Y}}$ (see equations (\ref{eq54}) and (\ref{eq55})) and when the Yukawa couplings are diagonalized, $\mathcal{Y}$ and $\tilde{\mathcal{Y}}$ are determined up to two phases \cite{hisano}. After the renormalization to the electroweak scale and dressing with a SUSY loop, we have:
\begin{equation}
\label{zaleznosc1}
C^{'abcd}(M_Z) \sim A_S \frac{m_{\tilde{w}}}{M_T m_{\tilde{q}}^2}  \frac{m_{u^a}(M_Z) m_{d^c}(M_Z) \delta^{ab}V_{CKM}^{*cd}}{M_W^2 \sin 2\beta_H}
\end{equation}
Thus, if the $LLLL$ operators are dominant\footnote{In case of large $\tan\beta_H$, the RRRR operators may become dominant. This does not spoil our argument, since in that case $\tau_p\sim (\tan\beta_H)^{-4}$ (\cite{goto}), so the proton lifetime is even shorter than in our estimation.}:
\begin{equation}
\label{tau1}
\tau_p\sim \frac{1}{\left( \tan\beta_H + \frac{1}{\tan\beta_H} \right)^2}
\end{equation}
and the proton lifetime is maximized for the models with small $\tan\beta_H$.

In many $\mathsf{SO(10)}$ models, the Yukawa couplings of the third generation of fermions result from the same coupling $\mathbf{16}_3\mathbf{16}_3\mathbf{10_H}$, which makes them all equal at the unification scale (see e.g. \cite{raby}). Since the top-bottom-tau unification requires large $\tan\beta_H$, the eq. (\ref{tau1}) yields that the proton lifetime is additionally suppressed in these models.

There is also a class of models, in which the light doublet $H_d$ is a mixture of states coming from the representations $\mathbf{10_H}$ and $\mathbf{16_H}$ as in (\ref{mixture}). In these models (e.g. \cite{albright}, \cite{babu}):
\begin{equation}
\frac{y_t}{y_b} = \frac{\tan\beta_H}{\cos\gamma_d}
\end{equation}
Thus $\tan\beta_H$ can be small, provided that $\cos\gamma_d\ll 1$. Assuming that the Wilson coefficients exhibit a hierarchy similar to that of the Yukawa couplings, we expect that the equation (\ref{zaleznosc1}) becomes:
\begin{equation}
C^{'abcd}(M_Z) \sim A_S \frac{m_{\tilde{w}}}{M_T m_{\tilde{q}}^2} \frac{ \mathrm{max}\{ m_{u^a}, m_{u^b}\}\mathrm{max}\{ m_{d^c}, m_{d^d}\} V^{ab}_{CKM} V^{*cd}_{CKM}}{M_W^2 \sin 2\beta_H \cos\gamma_d}
\end{equation}
(where all the quark masses are the running masses at $M_Z$) and the smallness of $\cos\gamma_d$ additionally suppresses the proton lifetime:
\begin{equation}
\label{suppression2}
\tau_p\sim \frac{\cos^2\gamma_d}{\left( \tan\beta_H + \frac{1}{\tan\beta_H} \right)^2}
\end{equation}
This suggests another way of raising the predictions for the proton lifetime. In a model with small $\tan\beta_H$, if we add the operators which make the Yukawa coupling of the top quark much larger than the other Yukawa couplings and do not contribute to proton decay, we could avoid the suppression (\ref{suppression2}). Such a model has been proposed in \cite{dvali}, but it reproduces the bad mass relations of the minimal $\mathsf{SU(5)}$.

\begin{figure}
\begin{center}
\rotatebox{0}{
\epsfig{file=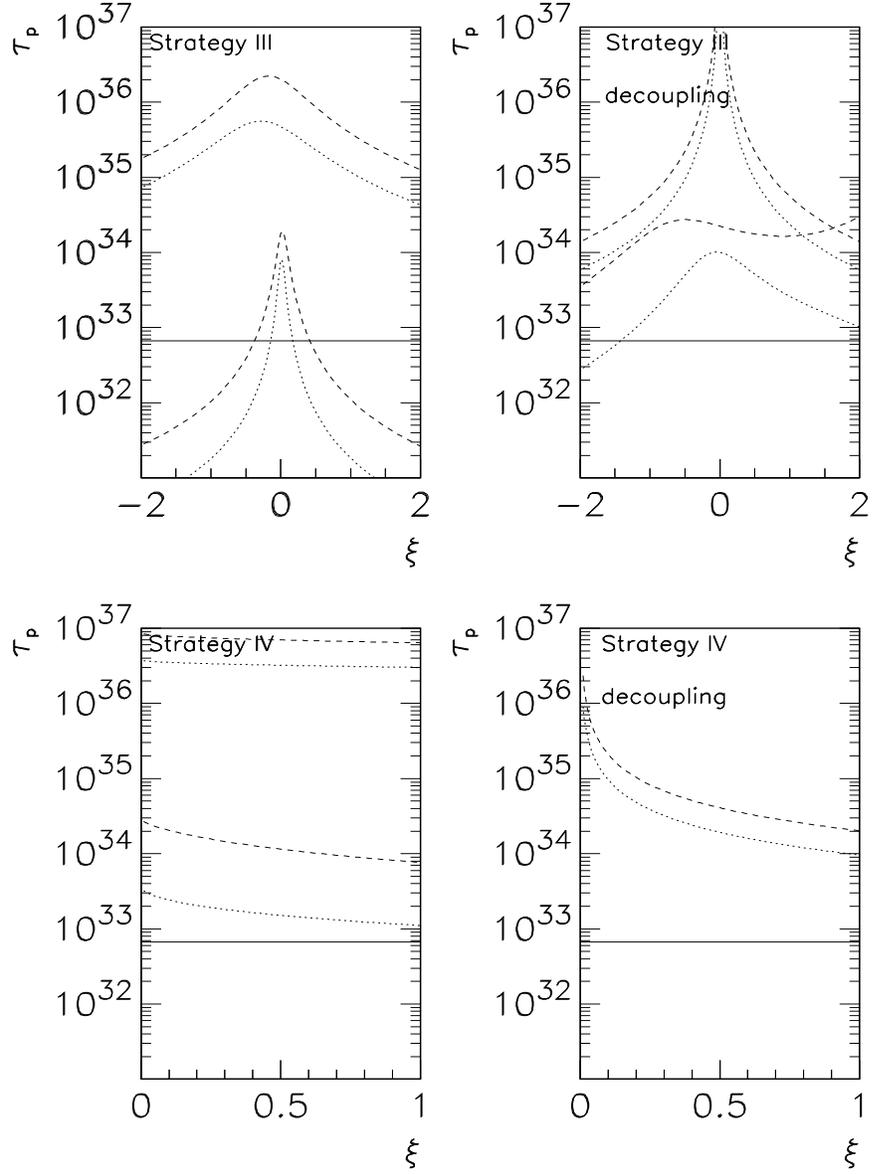,height=20cm,width=15cm}}
\caption{The proton lifetime (in years) in the dominant channels as a function of $\xi$ for Strategies I and II. See the caption of Figure (\ref{figa5}) and Sections \ref{s44} and \ref{s5} for the explanation.\label{figa4}}
\end{center}
\end{figure}

Despite this disadvantage, we analyzed a combination of the two models presented in \cite{dvali}. We assumed that the top quark is a mixture of the states coming from the spinor and adjoint representations:
\begin{equation}
t = \xi u_{\mathbf{16}_3} + \sqrt{1-\xi^2} u_\mathbf{45}
\end{equation}
Then the large mass of the top quark is generated by the operator $\mathbf{45}^2$ together with the operators (\ref{bcztery})--(\ref{bosiem})\footnote{It should be stressed that this model is quite different from the one discussed in Section 2, since all the light Higgs doublets reside solely in the spinor representations $\mathbf{16}_H+\mathbf{\bar{16}_H}$.}. Moreover, only $\xi y_t$ instead of $y_t$ enters the relevant contributions to the proton decay amplitude. We also assumed that the large masses of the colour triplets are generated by the following part of the superpotential:
\begin{equation}
W_T = b_1 \mathbf{16_H} \langle \mathbf{45_R} \rangle \mathbf{\bar{16}'_H} + b_2 \mathbf{\bar{16}_H} \langle \mathbf{45_R} \rangle \mathbf{16'_H} + b_3 \mathbf{16'_H} \langle \mathbf{1} \rangle \mathbf{\bar{16'}_H}
\end{equation} 
where $\mathbf{16'_H}+\mathbf{\bar{16}'_H}$ are fields which neither couple to matter nor acquire vevs. Then the effective mass of the colour triplets:
\begin{equation}
M_T = \frac{b_1b_2\langle \mathbf{45_R} \rangle^2 }{b_3\langle \mathbf{1} \rangle}
\end{equation}
can naturally be made larger than the unification scale $M_{GUT}$. Therefore, we set $M_T=2\cdot 10^{18}\mathrm{GeV}$ as in the other $\mathsf{SO(10)}$ models. The predictions for the proton lifetime as a function of $\xi$ are presented on Figure \ref{figa4}.

In all the cases, the predictions for the proton lifetime are consistent with the experimental data.

\section{Conclusions.}
\setcounter{equation}{0}

We have analyzed a class of SUSY $\mathsf{SO(10)}$ GUT's which can provide fermions with realistic masses. We have shown there is a lot of freedom in the parameters of such theories which does not affect the predictions for the low energy phenomena except for proton decay. We have proposed four strategies to increase the predicted proton lifetime. Though our solutions are clearly lacking naturalness, they show that no definite quantitative results can be obtained for the proton decay induced by the $d=5$, $\Delta B\neq 0$ operators. Therefore, only the predictions for proton decay induced by the heavy gauge bosons exchange, which are model independent, can be an unambiguous test of SUSY GUT's.

\section*{Acknowledgements.}

K.T. would like to thank Professor Stefan Pokorski for invaluable inspiracy and encouragement which made the completion of this work possible. He is also indebted to Piotr H. Chankowski for numerous discussions and allowing to use his computer codes. K.T. was supported partially by the EC Contract HPRN-CT-2000-00148 for years 2000-2004 and by the Polish State Committee for Scientific Research grant 5 P03B 119 20 for years 2001-2002.


\end{fmffile}
\end{document}